# Spectral Element Method for Vector Radiative Transfer Equation

J. M. Zhao[a], L. H. Liu[a,c]*, P. -f. Hsu[b], J. Y. Tan[c]

[a] *School of Energy Science and Engineering, Harbin Institute of Technology, 92 West Dazhi Street, Harbin 150001, People's Republic of China*

[b] *Department of Mechanical and Aerospace Engineering, Florida Institute of Technology, Melbourne, FL 32901, USA*

[c] *School of Automobile Engineering, Harbin Institute of Technology at Weihai, 2 West Wenhua Road, Weihai 264209, People's Republic of China*

**Abstract**

A spectral element method (SEM) is developed to solve polarized radiative transfer in multidimensional participating medium. The angular discretization is based on the discrete-ordinates approach, and the spatial discretization is conducted by spectral element approach. Chebyshev polynomial is used to build basis function on each element. Four various test problems are taken as examples to verify the performance of the SEM. The effectiveness of the SEM is demonstrated. The *h* and the *p* convergence characteristics of the SEM are studied. The convergence rate of *p*-refinement follows the exponential decay trend and is superior to that of *h*-refinement. The accuracy and efficiency of the higher order approximation in the SEM is well demonstrated for the solution of the VRTE. The predicted angular distribution of brightness temperature and Stokes vector by the SEM agree very well with the benchmark solutions in references. Numerical results show that the SEM is accurate, flexible and effective to solve multidimensional polarized radiative transfer problems.

*Keywords:* Spectral element method; Polarization; Vector radiative transfer

---

*Corresponding author. Tel.: +86-451-86402237; fax: +86-451-86221048.

*E-mail address:* lhliu@hit.edu.cn (L. H. Liu).

**Nomenclature**

| | |
|---|---|
| $\bar{\bar{\mathbf{D}}}$ | Matrix defined by Eq. (16b) |
| $\bar{\bar{\mathbf{E}}}$ | Identity matrix |
| $G$ | Incident radiation, W/m$^2$ |
| $H$ | Height or thickness, m |
| $\mathbf{i}, \mathbf{j}, \mathbf{k}$ | Unit vector of $x$-, $y$- and $z$-direction |
| $I_b$ | Black body radiative intensity, W/(m$^2$ μm sr) |
| $\mathbf{I}$ | Stokes vector $\mathbf{I} = (I, Q, U, V)^T$, W/(m$^2$ μm sr) |
| $\bar{\mathbf{I}}$ | Solution matrix of Stokes vector, defined by Eq. (16a) |
| $\bar{\bar{\mathbf{J}}}$ | Matrix defined by Eq. (16e) |
| $\bar{\bar{\mathbf{K}}}$ | Matrix defined by Eq. (16d) |
| $m$ | Relative refractive index |
| $M$ | Number of discrete ordinate directions |
| $\bar{\bar{\mathbf{M}}}$ | Matrix defined by Eq. (16c) |
| $\mathbf{n}$ | Unit normal vector |
| $N_{el}$ | Total number of elements |
| $N_{sol}$ | Total number of solution nodes |
| $N_\theta, N_\varphi$ | Number of zenith and azimuthal subdivisions |
| $p$ | Order of polynomial expansion |
| $P_l$ | Legendre polynomial of order $l$ |
| $\bar{\bar{\mathbf{P}}}$ | Reference scattering phase matrix |
| $\mathbf{r}$ | Spatial coordinates vector, m |
| $\bar{\bar{\mathbf{R}}}$ | Reflection matrix |
| **S** | Source term function defined by Eq. (11) |
| $\overline{\overline{\mathbf{S}}}$ | Matrix defined by Eq. (16f) |
| $T$ | Temperature, K |
| $T_B$ | Brightness temperature, K |
| $w$ | Weight of discrete ordinates approximation |
| $x, y, z$ | Cartesian coordinates |
| $\overline{\overline{\mathbf{Z}}}$ | Scattering phase matrix |
| $\varepsilon$ | Unpolarized emissivity |
| $\boldsymbol{\varepsilon}$ | Wall emission vector |
| $\Phi$ | Components of the reference scattering phase matrix $\overline{\overline{\mathbf{P}}}$ |
| $\varphi$ | Azimuthal angle |
| $\Gamma$ | Nodal basis function of spectral element method |
| $\eta$ | Direction cosine of y direction, $\eta = \sin\theta \sin\varphi$ |
| $\boldsymbol{\kappa}_a$ | Absorption vector, m$^{-1}$ |
| $\overline{\overline{\boldsymbol{\kappa}}}_e$ | Extinction matrix, m$^{-1}$ |
| $\kappa_e$ | Extinction coefficient, m$^{-1}$ |
| $\lambda$ | Wave length, μm |
| $\mu$ | Direction cosine of z direction, $\mu = \cos\theta$ |
| $\theta$ | Zenith angle |
| $\theta_i$ | Incident angle of Fresnel reflection |
| $\rho$ | Unpolarized reflectivity |
| $\sigma$ | Stefan-Boltzmann constant, W/(m$^2$K$^4$) |
| $\tau_L$ | Optical thickness |

| $\Theta$ | Angle between incident and scattered radiation |
|---|---|
| $\boldsymbol{\Omega}, \boldsymbol{\Omega}'$ | Unit vector of radiation direction, $\boldsymbol{\Omega} = \xi\mathbf{i} + \eta\mathbf{j} + \mu\mathbf{k}$ |
| $\Omega$ | Solid angle |
| $\xi$ | Direction cosine of x direction, $\xi = \sin\theta\cos\varphi$ |
| $\psi, \psi'$ | Rotation angle, defined in Fig. 1 |
| $\bar{\bar{\mathfrak{R}}}$ | Rotation matrix of Stokes vector |

*Subscripts*

| $Fr$ | Fresnel surface |
|---|---|
| $i, j$ | Spatial node index |
| $k$ | Index of Stokes parameters |
| $La$ | Lambertian surface |
| $w$ | Wall quantity |

*Superscripts*

| $n$ | The *m*-th discrete ordinate direction |
|---|---|

## 1. Introduction

Polarization is one of the most important characteristics of light. The interaction of light with medium such as scattering by particles and reflection at the boundary will generally alter the polarization of an incident light. Polarized radiative transfer has found wide applications in the fields of remote sensing [1-3], atmospheric optics and so on. The vector radiative transfer theory is the theoretical basis of the polarized radiative transfer, which is based on the vector radiative transfer equation of four Stokes parameters. By comparison with the scalar radiative transfer equation, the vector radiative transfer equation (VRTE) can accurately account for the polarization of light during transport in participating medium, while the solution of VRTE is more complicate for it is composed of four coupled equations of Stokes parameters, which partially limited its broad application.

Accurate and computationally efficient radiative transfer calculations are important for many applications, such as the retrieval of atmospheric and oceanic constituents from remotely sensed satellite observations. Several numerical methods have been proposed for the solution of the VRTE, such as $F_N$ method [4], doubling-adding method [5], discrete ordinate method [6-8], Monte Carlo methods [9, 10], and spectral methods [11, 12]. The method based on discrete ordinates approach for the solution of radiative transfer equation has advantages such as being easy and efficient to treat multidimensional problems, and also flexible to deal with complex media and boundary property. However, most of the already developed discrete ordinates based methods is often of lower order accuracy (such as, 1st order or 2nd order) and only offer the $h$-refinement scheme, i.e. by reducing the element size $h$ to obtain convergent solution, as a result, re-meshing are often required in order to gain the wanted accuracy. Though the spectral methods can give high order accuracy, they are often limited only to simple geometry due to their global spectral basis functions are difficult to be built on a general computational domain.

Spectral element method was originally proposed by Patera [13] for the solution of fluid flow problem, which has found successful applications in many disciplines [13-16] in recent years. There are basically two types of refinement scheme provided by the spectral element method (SEM) to achieve convergence. The most common one is to reduce element size in the regions where higher resolution is required, which is called $h$-refinement in literature. The second type refinement is the $p$-refinement, in which the number of elements and their sizes are fixed but the approximation order inside the elements is increased. As a result, SEM is very flexible for the solution of multidimensional problem in complex geometries with higher order accuracy. Recently, Zhao and Liu [17, 18] introduced the SEM to solve multidimensional scalar radiative transfer problems. Their numerical results demonstrated that higher order spectral approximation on each element is effective and efficient to solve multidimensional radiative transfer problems.

In present work, the spectral element method is extended to solve the multidimensional VRTE. The angular discretization is based on the discrete-ordinates approach, and the spatial discretization is conducted by spectral element approach. Four various test cases of polarized radiative transfer are taken to verify the performance of the method.

## 2. Mathematical Formulation

### 2.1. Three-Dimensional Vector Radiative Transfer Equation

Based on the incoherent addition principle of Stokes parameters, the steady state three-dimensional VRTE for polarized monochromatic radiation in medium with randomly-oriented particles can be written as [19, 20]

$$\mathbf{\Omega} \cdot \nabla \mathbf{I}(\mathbf{r}, \mathbf{\Omega}) + \overline{\overline{\mathbf{\kappa}}}_e \mathbf{I}(\mathbf{r}, \mathbf{\Omega}) = \mathbf{\kappa}_a I_b(\mathbf{r}) + \int_{4\pi} \overline{\overline{\mathbf{Z}}}(\mathbf{r}, \mathbf{\Omega}' \to \mathbf{\Omega}) \mathbf{I}(\mathbf{r}, \mathbf{\Omega}') d\mathbf{\Omega}' \qquad (1)$$

Where $\mathbf{I} = (I, Q, U, V)^T$ is the vector of Stokes parameters; $\mathbf{\Omega}$ is the discrete direction vector, $\mathbf{r}$ is the spatial coordinate vector; $\overline{\overline{\mathbf{\kappa}}}_e$ is the extinction matrix; $\mathbf{\kappa}_a$ is the absorption vector and $\mathbf{\kappa}_a = (1, 0, 0, 0)^T$ for unpolarized emission; $I_b$ is the black body intensity; $\overline{\overline{\mathbf{Z}}}$ is the scattering phase matrix. The geometry for the definition of polarization reference frame used in Eq. (1) is shown in Fig. 1. For a scatter at the origin, the scattering plane is $OPP'$, the meridian plane for the incident ray $\mathbf{\Omega}'$ and scattered outgoing ray $\mathbf{\Omega}$ are $OZP'$ and $OZP$, respectively. The parallel ('**l**') and perpendicular ('**r**') components of the polarized beam are parallel and perpendicular to the meridian plane, respectively. Because the definition of the polarization component of the incident and scattered beam is not aligned with the scattering plane $OP'P$, rotation of Stokes parameters is needed in calculating the scattering phase matrix $\overline{\overline{\mathbf{Z}}}$. The transform of the scattering matrix $\overline{\overline{\mathbf{P}}}$ to phase matrix $\overline{\overline{\mathbf{Z}}}$ is expressed as [19]

$$\overline{\overline{\mathbf{Z}}}(\mathbf{r}, \mathbf{\Omega}' \to \mathbf{\Omega}) = \overline{\overline{\mathfrak{R}}}(\pi - \psi) \overline{\overline{\mathbf{P}}}(\mathbf{r}, \mathbf{\Omega}' \to \mathbf{\Omega}) \overline{\overline{\mathfrak{R}}}(-\psi') \qquad (2)$$

here $\psi, \psi' \in [0, \pi]$ are the rotation angles shown in Fig. 1, $\overline{\overline{\mathfrak{R}}}(\psi)$ is the rotation matrix given as

$$\overline{\overline{\mathfrak{R}}}(\psi) = \begin{bmatrix} 1 & 0 & 0 & 0 \\ 0 & \cos 2\psi & \sin 2\psi & 0 \\ 0 & -\sin 2\psi & \cos 2\psi & 0 \\ 0 & 0 & 0 & 1 \end{bmatrix} \quad (3)$$

To complete the transform given in Eq. (2), $\cos\psi$ and $\cos\psi'$ are needed to be obtained. Here a vector notation is presented to calculate these two cosines of the rotation angles. Details are given in the Appendix Section A. Without loss of generality, only the phase matrix contains six independent nonzero elements [21] is considered, which is for randomly oriented particles with a plane of symmetry. As for the convenience of program implementation, the scattering phase matrix $\overline{\overline{Z}}$ [Eq.(2)] can be explicitly expanded out as given in Ref. [5], and each element of the scattering matrix $\overline{\overline{P}}$ is expanded by the Legendre polynomials following the approach of Ref. [5].

In general, the boundary condition for the VRTE [Eq.(1)] which takes into account Lambertian (diffuse) emission and reflection, Fresnel emission reflection, and collimated beam irradiation can be written as [8]

$$\begin{aligned} \mathbf{I}(\mathbf{r}_w, \mathbf{\Omega}) &= \left[ f_{Fr} \boldsymbol{\varepsilon}_{Fr} \left( |\mathbf{n}_w \cdot \mathbf{\Omega}| \right) + f_{La} \boldsymbol{\varepsilon}_{La} \right] I_b(\mathbf{r}_w) + \mathbf{I}_C(\mathbf{r}_w) \delta(\mathbf{\Omega} - \mathbf{\Omega}_C) \\ &+ f_{Fr} \overline{\overline{\mathbf{R}}}_{Fr} \left( |\mathbf{n}_w \cdot \mathbf{\Omega}| \right) \mathbf{I}(\mathbf{r}_w, \mathbf{\Omega}'') \\ &+ \frac{f_{La}}{\pi} \left[ \int_{\mathbf{n}_w \cdot \mathbf{\Omega}' > 0} \overline{\overline{\mathbf{R}}}_{La} \mathbf{I}(\mathbf{r}_w, \mathbf{\Omega}') |\mathbf{\Omega}' \cdot \mathbf{n}_w| d\Omega' \right] \end{aligned} , \quad \mathbf{\Omega} \cdot \mathbf{n}_w < 0 \quad (4)$$

where, $\boldsymbol{\varepsilon}_{Fr}$ and $\boldsymbol{\varepsilon}_{La}$ are the surface emissivity vector of the Fresnel and Lambertian emission, respectively; $f_{Fr}$ and $f_{La} = (1 - f_{Fr})$ denote the fraction of Fresnel reflection and diffuse reflection, respectively; $\mathbf{I}_C$ is the irradiation flux vector of the collimated beam; $\mathbf{\Omega}''$ is the corresponding incident direction of the current reflected beam of direction $\mathbf{\Omega}$; $\mathbf{n}_w$ is the unit normal vector of the boundary; $\overline{\overline{\mathbf{R}}}_{Fr}$ is the reflection matrix defined by the Fresnel's law, given as [1]

$$\overline{\overline{\mathbf{R}}}_{Fr}(\cos\theta_i) = \begin{bmatrix} (|r_v|^2 + |r_h|^2)/2 & (|r_v|^2 - |r_h|^2)/2 & 0 & 0 \\ (|r_v|^2 - |r_h|^2)/2 & (|r_v|^2 + |r_h|^2)/2 & 0 & 0 \\ 0 & 0 & \mathrm{Re}(r_v r_h^*) & -\mathrm{Im}(r_v r_h^*) \\ 0 & 0 & \mathrm{Im}(r_v r_h^*) & \mathrm{Re}(r_v r_h^*) \end{bmatrix} \quad (5)$$

in which

$$r_v = \frac{\cos\theta_i - \sqrt{m^2 + \cos^2\theta_i - 1}}{\cos\theta_i + \sqrt{m^2 + \cos^2\theta_i - 1}}, \quad r_h = \frac{m^2\cos\theta_i - \sqrt{m^2 + \cos^2\theta_i - 1}}{m^2\cos\theta_i + \sqrt{m^2 + \cos^2\theta_i - 1}} \tag{6}$$

Here $\theta_i$ is incident angle, $m$ is relative refractive index of the substrate, and the corresponding emissivity vector $\mathbf{\varepsilon}_{Fr}$ for the Fresnel surface is

$$\mathbf{\varepsilon}_{Fr}(\cos\theta_i) = \begin{bmatrix} 1 - (|r_v|^2 + |r_h|^2)/2 \\ (|r_v|^2 - |r_h|^2)/2 \\ 0 \\ 0 \end{bmatrix} \tag{7}$$

$\overline{\overline{\mathbf{R}}}_{La}$ and $\mathbf{\varepsilon}_{La}$ are the reflection matrix and emissivity vector of the Lambertian surface, respectively,

$$\overline{\overline{\mathbf{R}}}_{La} = \begin{bmatrix} \rho & 0 & 0 & 0 \\ 0 & 0 & 0 & 0 \\ 0 & 0 & 0 & 0 \\ 0 & 0 & 0 & 0 \end{bmatrix}, \quad \mathbf{\varepsilon}_{La} = \begin{bmatrix} \varepsilon \\ 0 \\ 0 \\ 0 \end{bmatrix} \tag{8}$$

where $\rho$ and $\varepsilon$ are the unpolarized reflectivity and emissivity of the boundary, respectively.

The boundary condition given in Eq. (4) assumes the reflection (or refraction) optical planes (planes contain reflected direction and the normal vector of the boundary) coincide with the reference planes of polarization (the meridian planes shown in Fig. 1), this is often the case in the one-dimensional radiative transfer problem, where the normal vector of the boundary surface coincides with z-direction (the direction used to define the reference planes of polarization), which results in the coincidence of the reference planes of polarization and the reflection optical planes. However, in multidimensional case, the normal vector of the boundary may not coincide with z-direction, and the optical planes are not always coincident with the reference planes of polarization. The original Fresnel reflection matrix [Eq. (5)] and emissivity vector [Eq. (7)] are defined originally in the optical plane. In this case, similar rotation of Stokes parameters is needed as that is done in the calculation of scattering phase matrix [Eq. (2)], the boundary condition Eq. (4) is rewritten as

$$\mathbf{I}(\mathbf{r}_w,\mathbf{\Omega}) = \left[ f_{Fr}\overline{\overline{\mathfrak{R}}}(\psi)\mathbf{\varepsilon}_{Fr}(|\mathbf{n}_w\cdot\mathbf{\Omega}|) + f_{La}\mathbf{\varepsilon}_{La} \right] I_b(\mathbf{r}_w) + \mathbf{I}_C(\mathbf{r}_w)\delta(\mathbf{\Omega}-\mathbf{\Omega}_C)$$
$$+ f_{Fr}\overline{\overline{\mathfrak{R}}}(\pi-\psi)\overline{\overline{\mathbf{R}}}_{Fr}(\mathbf{n}_w\cdot\mathbf{\Omega})\overline{\overline{\mathfrak{R}}}(-\psi'')\mathbf{I}(\mathbf{r}_w,\mathbf{\Omega}'') \quad , \quad \mathbf{\Omega}\cdot\mathbf{n}_w < 0 \quad (9)$$
$$+ \frac{f_D}{\pi}\left[ \int_{\bar{\mathbf{n}}_w\cdot\bar{\mathbf{\Omega}}'>0} \overline{\overline{\mathbf{R}}}_{La}\, \mathbf{I}(\mathbf{r}_w,\mathbf{\Omega}')|\mathbf{\Omega}'\cdot\mathbf{n}_w|d\Omega' \right]$$

Here, $\psi$ is the angle between the emission optical plane and the meridian plane of direction $\mathbf{\Omega}$, $\psi''$ is the angle between the optical plane and the meridian plane of the original incident direction of Fresnel reflection $\mathbf{\Omega}''$. By taking $\mathbf{\Omega}''$ as the incident direction $\mathbf{\Omega}'$ and $\mathbf{\Omega}$ as the scattered direction defined in the calculation of scattering phase matrix as shown in Fig. 1, the relation of rotation is the same as in the calculation of scattering phase matrix [Eq. (2)]. It is noted that the rotation of Stokes parameters is not needed for the reflection and emission of the Lambertian boundary because they are unpolarized. For $f_{Fr}$ = 1, Eq. (4) or Eq. (9) reduced to purely Fresnel boundary condition, or purely Lambertian boundary condition when $f_{La}$ = 0.

*2.2. Spectral Element Method Discretization*

In this section, a spectral element method is presented to solve the VRTE, Eq. (1), in which the angular discretization is conducted by discrete ordinates approach and the spatial discretization by spectral element approach. The discrete ordinate form of the VRTE can be written as

$$\mathbf{\Omega}^n\cdot\nabla\mathbf{I}(\mathbf{r},\mathbf{\Omega}^n) + \overline{\overline{\mathbf{\kappa}}}_e\mathbf{I}(\mathbf{r},\mathbf{\Omega}^n) = \mathbf{S}(\mathbf{r},\mathbf{\Omega}^n) \quad (10)$$

where source function $\mathbf{S}(\mathbf{r},\mathbf{\Omega}^n)$ is defined as

$$\mathbf{S}(\mathbf{r},\mathbf{\Omega}^n) = \mathbf{\kappa}_a I_b(\mathbf{r}) + \sum_{n'=1}^{M}\overline{\overline{\mathbf{Z}}}(\mathbf{r},\mathbf{\Omega}^{n'}\to\mathbf{\Omega}^n)\mathbf{I}(\mathbf{r},\mathbf{\Omega}^{n'})w^{n'} \quad (11)$$

Here $\mathbf{\Omega}^n$ or $\mathbf{\Omega}^{n'}$ is the discrete direction, and $w^{n'}$ is the corresponding weight. There are many discrete ordinate set available for the solution of radiative transfer equation [22] and each set can be used in Eq. (10).

Spectral element method can be considered as a special kind of finite element method. The feature of SEM is that the nodal basis functions are constructed on each element by orthogonal polynomial expansion. In present approach, the Chebyshev polynomial expansion is employed. The one-dimensional nodal basis functions are Lagrange interpolation polynomials through the Chebyshev-Gauss-Lobatto points. Multidimensional nodal basis functions on structured element (quadrilateral, cube) are constructed by tensor product from one-dimensional nodal basis functions. Details on building the nodal basis function were described in Ref. [17]. With the help of spectral element nodal basis function $\Gamma_i(\mathbf{r})$, the unknown Stokes vector $\mathbf{I}^n(\mathbf{r}) = \mathbf{I}(\mathbf{r}, \mathbf{\Omega}^n)$, $\overline{\overline{\mathbf{\kappa}}}_e(\mathbf{r})\mathbf{I}(\mathbf{r}, \mathbf{\Omega}^n)$ and $\mathbf{S}(\mathbf{r}, \mathbf{\Omega}^n)$ can be approximated as

$$\mathbf{I}^n(\mathbf{r}) \simeq \sum_{i=1}^{N_{sol}} \mathbf{I}_i^n \Gamma_i(\mathbf{r}), \quad \overline{\overline{\mathbf{\kappa}}}_e(\mathbf{r})\overline{\mathbf{I}}(\mathbf{r}, \mathbf{\Omega}^n) \simeq \sum_{i=1}^{N_{sol}} \overline{\overline{\mathbf{\kappa}}}_{e,i}\mathbf{I}_i^n \Gamma_i(\mathbf{r}), \quad \mathbf{S}^n(\mathbf{r}) \simeq \sum_{i=1}^{N_{sol}} \mathbf{S}_i^n \Gamma_i(\mathbf{r}) \tag{12}$$

where $\mathbf{I}_i^n$, $\overline{\overline{\mathbf{\kappa}}}_{e,i}$ and $\mathbf{S}_i^n$ denote the Stokes vector, extinction matrix and $\mathbf{S}(\mathbf{r}, \mathbf{\Omega}^n)$ at solution nodes $i$, respectively; and $N_{sol}$ is the total number of solution nodes. Substituting Eq. (12) into Eq. (10) and applying the standard Galerkin approach, namely, Eq. (10) is weighted by $\Gamma_j(\mathbf{r})$ and integrated over the solution domain, we obtain

$$\sum_{i=1}^{N_{sol}} \left[ <\mathbf{\Omega}^n \cdot \nabla\Gamma_i(\mathbf{r}), \Gamma_j(\mathbf{r})> + <\Gamma_i(\mathbf{r}), \Gamma_j(\mathbf{r})> \overline{\overline{\mathbf{\kappa}}}_{e,i} \right] \mathbf{I}_i^n = \sum_{i=1}^{N_{sol}} <\Gamma_i(\mathbf{r}), \Gamma_j(\mathbf{r})> \mathbf{S}_i^n \tag{13}$$

In which the inner product $<\bullet, \bullet>$ is defined as $<f, g> = \int_V f\, g\, \mathrm{d}V$. Equation (13) is further rearranged as

$$\sum_{i=1}^{N_{sol}} \left[ <\mathbf{\Omega}^n \cdot \nabla\Gamma_i(\mathbf{r}), \Gamma_j(\mathbf{r})> + <\Gamma_i(\mathbf{r}), \Gamma_j(\mathbf{r})> \kappa_{e,i} \right] \mathbf{I}_i^n$$
$$= \sum_{i=1}^{N_{sol}} <\Gamma_i(\mathbf{r}), \Gamma_j(\mathbf{r})> \left[ \kappa_{e,i}\overline{\overline{\mathbf{E}}} - \overline{\overline{\mathbf{\kappa}}}_{e,i} \right] \mathbf{I}_i^m + \sum_{i=1}^{N_{sol}} <\Gamma_i(\mathbf{r}), \Gamma_j(\mathbf{r})> \mathbf{S}_i^n \tag{14}$$

where $\kappa_e$ denotes the unpolarized extinction coefficient and $\overline{\overline{\mathbf{E}}}$ is an identity matrix. Equation (14) can be written in matrix form as

$$\left[ \overline{\overline{\mathbf{D}}}(\mathbf{\Omega}^n) + \overline{\overline{\mathbf{M}}} \bullet \overline{\overline{\mathbf{K}}} \right] \overline{\overline{\mathbf{I}}}(\mathbf{\Omega}^n) = \overline{\overline{\mathbf{M}}}\, \overline{\overline{\mathbf{J}}}(\mathbf{\Omega}^n) + \overline{\overline{\mathbf{M}}}\, \overline{\overline{\mathbf{S}}}(\mathbf{\Omega}^n) \tag{15}$$

here '$\bullet$' denotes the Hadamard product (element-wise matrix multiplication), and the matrices are defined as follows

$$\overline{\overline{\mathbf{I}}}(\mathbf{\Omega}^n) = \left[ I_{ik}^n \right]_{i=1,N_{sol};k=1,4} = \left[ \left( I_i^n, Q_i^n, U_i^n, V_i^n \right) \right]_{i=1,N_{sol}} \tag{16a}$$

$$\overline{\overline{\mathbf{D}}}(\mathbf{\Omega}^n) = \left[ D_{ji}(\mathbf{\Omega}^n) \right]_{j=1,N_{sol};i=1,N_{sol}} = \left[ <\mathbf{\Omega}^n \cdot \nabla\Gamma_i(\mathbf{r}), \Gamma_j(\mathbf{r})> \right]_{j=1,N_{sol};i=1,N_{sol}} \tag{16b}$$

$$\overline{\overline{\mathbf{M}}} = \left[ M_{ji} \right]_{j=1,N_{sol};i=1,N_{sol}} = \left[ <\Gamma_i(\mathbf{r}), \Gamma_j(\mathbf{r})> \right]_{j=1,N_{sol};i=1,N_{sol}} \tag{16c}$$

$$\overline{\overline{\mathbf{K}}} = \left[ K_{ji} \right]_{j=1,N_{sol};i=1,N_{sol}} = \left[ \kappa_{e,i} \right]_{j=1,N_{sol};i=1,N_{sol}} \quad (16d)$$

$$\overline{\overline{\mathbf{J}}}(\mathbf{\Omega}^n) = \left[ J_{ik}(\mathbf{\Omega}^n) \right]_{i=1,N_{sol};k=1,4} = \left[ \kappa_{e,i}(\mathbf{I}_i^n)^T - (\overline{\overline{\kappa}}_{e,i}\ \mathbf{I}_i^n)^T \right]_{i=1,N_{sol}} \quad (16e)$$

$$\overline{\overline{\mathbf{S}}}(\mathbf{\Omega}^n) = \left[ S_{ik}(\mathbf{\Omega}^n) \right]_{i=1,N_{sol};k=1,4} = \left\{ \boldsymbol{\kappa}_{a,i}^T I_{b,i} + \sum_{n'=1}^{M} \left[ \overline{\overline{\mathbf{Z}}}_i(\mathbf{\Omega}^{n'} \rightarrow \mathbf{\Omega}^n) \mathbf{I}_i^{n'} \right]^T w^{n'} \right\}_{i=1,N_{sol}} \quad (16f)$$

The unknown field of Stokes vector can be obtained by solving Eq. (15) with $\overline{\overline{\mathbf{I}}}(\mathbf{\Omega}^m)$ as the unknown variable.

*2.4. Solution Procedure*

The matrix equations Eq. (15) are solved on each discrete ordinate direction, and the boundary conditions are imposed by collocation approach. Because the right hand side of Eq. (15) in direction $m$ contains the Stokes vector of other directions, global iteration are necessary to update right hand side term. The implementation of the SEM can be carried out according to the following routine:

Step 1. Mesh the solution domain with non-overlapping elements through conventional mesh tools for FEM or FVM method.

Step 2. Choose the order of Chebyshev polynomial to build elemental basis function and generate the solution nodes for each element with Gauss-Chebyshev-Lobatto points.

Step 3. Build the basis function for each element, and integrate to get the stiffness matrices defined in Eq. (16b) and (16c) according to the precalculation approach of stiffness matrices described in Ref. [17].

Step 4. Begin the global iteration to update the right hand side term.

Step 5. Begin loop each angular direction for $n = 1,...,M$, and calculate corresponding right hand side matrices $\overline{\overline{\mathbf{J}}}(\mathbf{\Omega}^n)$, $\overline{\overline{\mathbf{S}}}(\mathbf{\Omega}^n)$.

Step 6. Impose boundary condition according to the collocation approach.

Step 7. Solve the linear Eq. (15) to obtain the Stokes vector on each solution nodes along the angular direction $m$. Return to Step 5 for the next angular direction.

Step 8.   Terminate the iteration process if the stop criterion is satisfied. Otherwise go back to Step 4.

In this paper, the maximum relative error $10^{-4}$ of incident radiation $G(\mathbf{r}) = \int_{4\pi} I(\mathbf{r},\mathbf{\Omega})\mathrm{d}\Omega$ ($\|G_{new} - G_{old}\| / \|G_{new}\|$) is taken as the stop criterion for the global iteration.

## 3. Results and Discussion

To verify the performance of the SEM presented in this paper, four test cases are selected. The first three cases are one-dimensional problems, while the last one is a two-dimensional problem. For the sake of quantitative comparison to benchmark results, the relative error of the SEM solution is defined as

$$\text{Relative Error \%} = \frac{\int |\text{SEM solution} - \text{Benchmark result}| \mathrm{d}x}{\int |\text{Benchmark result}| \mathrm{d}x} \times 100 \qquad (17)$$

### 3.1. Case 1: Isothermal non-scattering absorbing atmosphere

The optical thickness based on the thickness $H$ of the atmosphere is 0.08. Below the atmosphere is sea, of which the refractive index is 3.724–2.212$i$. Temperatures of the water surface and the atmosphere are 300 K. The upper surface of the atmosphere is considered nonemitting. For this case, it is convenient to apply the ray tracing technique to obtain a precise analytical solution (see Appendix Section B). The SEM is applied to obtain the angular distribution of brightness temperature of Stokes parameters $I$ ($T_{B,I}$) and $Q$ ($T_{B,Q}$) at frequency 85.5 Ghz. The brightness temperature is calculated based on the Rayleigh-Jeans Law as

$$T_{B,I} = \frac{C_2}{2C_1}\lambda^4 I, \quad T_{B,Q} = \frac{C_2}{2C_1}\lambda^4 Q \qquad (18)$$

Where $C_1 = 0.59544 \times 10^8 \text{ W}\mu\text{m}^4/\text{m}^2$ and $C_2 = 1.4388 \times 10^4 \text{ }\mu\text{m K}$.

Figure 2(a) and 2(b) present the solved angular distributions of $T_{B,I}$ and $T_{B,Q}$, respectively, at three different positions: $z/H$ = 0, 0.5, and 1. The results obtained with ray tracing technique are also shown and taken as benchmarks. Noticing the axisymmetry of this problem, only the zenith angle is subdivided. In these solutions of the SEM, the angular discretization uses piecewise constant approximation (PCA) [18, 23] with $N_\theta = 40$, and the layer of atmosphere is subdivided into 6 elements and 4$^{th}$ order polynomial approximation is used. It is seen that the result obtained with the SEM agree very well with the benchmark result of ray tracing method for different component of polarization and at different position. The maximum relative error is less than 1.2%. A peak is observed at about $\mu = 0.3$ in the angular distribution of $T_{B,Q}$ for different positions, which is due to the polarization ability of the water surface having a maximum. For upward direction, $\mu > 0$, $T_{B,Q}$ decreases with the increasing of distance to the sea surface for different positions due to absorption of the atmosphere.

The effects of the polynomial order $p$ and total number of elements (namely, reducing the element size) on the convergence characteristics of the SEM is further studied. For convenience of interpreting the accuracy of the method, here the relative error is evaluated based on the solution of the incident radiation $G$, and the angular integration of $Q$, $G_Q$, at each spatial position, i.e.

$$G_Q(\mathbf{r}) = \int_{4\pi} Q(\mathbf{r}, \Omega) d\Omega \tag{19}$$

The relative errors of the SEM for the solution of $G$ and $G_Q$ at three different atmosphere optical thickness, namely, $\tau_L = 0.08$, 1 and 5, are presented in Fig. 3(a) and 3(b) for various spatial discretization schemes. The horizontal axis is given as the total number of solution nodes $N_{sol} = pN_{el} + 1$, which will determine the size of stiffness matrix and hence is a measure of computational cost. The *h*-refinement test is conducted for which the polynomial order is a fixed value of $p = 2$. While for the *p*-refinement test, the number of elements is a fixed number of $N_{el} = 2$. As can be seen from Fig. 3, the relative errors decay very fast with increasing polynomial order $p$ and approximately follow the exponential trend for different atmosphere optical thickness, while the convergence rate when increasing the number of elements is relatively very slow for the solutions of $G$ and $G_Q$. The effect of variable order of spectral approximation on the convergence characteristics of *h*-refinement is shown in Fig. 4 for the solution of $G$. Generally, the convergence rate of *h*-refinement under different order of spectral approximation is slow than the *p*-refinement. However, the convergence rate of *h*-refinement improves significantly with increasing the spectral approximation order $p$. This well demonstrates the accuracy and efficiency of higher order approximation for solving the VRTE. The SEM is capable of both *h*-refinement and *p*-refinement, which is flexible and effective to solve the polarized radiative transfer.

*3.3. Case 2: Mie scattering of collimated beam*

In this case, the '*L*=13 problem' of Garcia and Siewert [4] is considered. As seen in Fig. 5, an oblique solar beam (oblique angle is $\cos\theta_c = 0.2$, and $\varphi_c = \pi/2$) penetrates into an scattering atmosphere from above. The irradiation flux intensity of the beam is $\pi$. The optical thickness based on the thickness of the atmosphere *H* is 1, and the single scattering albedo is 0.99. The phase matrix is for Mie scattering at a wavelength of 0.951 μm from a gamma distribution of particles with effective radius of 0.2 μm, effective variance of 0.07, and refractive index *n*=1.44. The Legendre coefficients for the four unique elements of the phase matrix have been obtained by Evens and Stephen [5]. The lower boundary of the atmosphere is diffuse reflection with $\rho$=0.1. The emission of the atmosphere and the boundary are omitted. The SEM is applied to obtain the angular distribution of Stokes parameters along $\mu$ at plane $\varphi$=0. Figure 6(a), 6(b), and 6(c) present the angular distribution of the four Stokes parameters solved by the SEM at three different positions, $z/H$ =0, 0.5 and 1, compared with the results obtained using the $F_N$ method by Garcia and Siewert [4]. Here, $I_*$, $Q_*$, $U_*$, and $V_*$ denotes the diffuse component of the Stokes vector, this follows the notation used in Ref. [4]. Two angular discretization scheme is selected, namely, PCA schemes with $N_\theta \times N_\varphi = 20\times 40$ and $N_\theta \times N_\varphi = 40\times 40$. The atmosphere is subdivided into 6 elements and 4$^{\text{th}}$ order polynomial approximation is used. At different position, the Stokes parameters obtained with the SEM agree very well with the results obtained using by Garcia and Siewert [4] using the $F_N$ method. The selected angular discretization is demonstrated to be enough to obtain convergent result. The maximum relative error is less than 3%, and no observable difference can be seen from the figure. For $z/H = 0$, which is the start surface of the collimated beam transmit into the medium and the beam has not yet been scattered, hence the diffuse components of Stokes vector are zero for downward direction ($\mu > 0$). For upward direction ($\mu < 0$), due to multiple scattering and reflection of the bottom surface, the original unpolarized beam is polarized. For $z/H = 0.5$, all components of Stokes vector are nonzero for both upward and downward directions due to multiple scattering. For $z/H = 1$, $Q_*$, $U_*$, and $V_*$ components are zero for upward direction ($\mu < 0$), which is due to the depolarization of the Lambertian bottom surface.

The effects of the polynomial order $p$ and total number of elements on the convergence characteristics of the SEM is further studied for this scattering case. The convergence characteristics of the $h$- and $p$- refinement of the SEM for the solution of incident radiation $G$ and the angular integration of $Q(G_Q)$ is presented in Fig. 7. The $h$-refinement test is conducted for which the polynomial order takes a fixed value of $p = 2$, and the $p$-refinement test is conducted with a fixed number of elements $N_{el} = 2$. It is seen that the convergence rate of $p$- refinement follows the exponential law, while the convergence rate of $h$-refinement is far slower, for solving $G$ and $G_Q$, which follows the same trend as previous case. This demonstrates the SEM is also accurate, flexible and effective to solve the polarized radiative transfer in a scattering media.

### 3.2. Case 3: Two-layer scattering atmosphere

A two-layer atmosphere above the sea is considered. The lower layer of this atmosphere is modeled as rain and the upper layer is taken as ice particles. The thicknesses of the atmosphere is $H = 8$ km. The extinction coefficient of the upper atmosphere ($z = 4$ km ~ 8 km) is 0.13536 km$^{-1}$ and with a scattering albedo of 0.98719. The extinction coefficient of the lower atmosphere ($z = 0$ km ~ 4 km) is 0.15224 km$^{-1}$ with a scattering albedo of 0.38175. The Legendre series expansion coefficients of the scattering phase matrix of the rain layer and the ice layer are presented in Ref. [5]. The temperature of the atmosphere linearly decreases with the height from $z = 0$ km increases to 8 km and the temperatures of the upper surface of the atmosphere and sea surface are 2.7 K and 300 K, respectively. The refractive index of the water surface is 3.724–2.212$i$. Evans and Stephens [5] studied this case using the doubling-adding method. Here, SEM is applied to solve the brightness temperature $T_{B,I}$ and $T_{B,Q}$ at frequency 85.5 Ghz. The angular discretization uses PCA scheme [18, 23] with $N_\theta \times N_\varphi = 40 \times 40$, and the atmosphere is subdivided into 6 elements and 4$^{th}$ order polynomial approximation is used. Figure 8(a) and 8(b) present the angular distributions of brightness temperature $T_{B,I}$ and $T_{B,Q}$ solved by the SEM, respectively, at position $z/H = 0$ and $z/H = 1$, and compared with the results obtained by Evans and Stephens [5]. The results obtained with the SEM agree very well with the result obtained with the doubling-adding method, no observable difference can be seen from these figures. Similar to previous case, due to the polarization ability of the water surface has a maximum, for upward direction, $\mu > 0$, a peak is observed in the angular distribution of $T_{B,Q}$ for different positions, however, the angular position of the peaks for different height is significantly altered by multiple scattering of medium, such as, for $z/H = 0$, the angular position of the peak is about $\mu = 0.4$, while for $z/H = 1$ the angular position of the peak move to about $\mu = 0.6$.

Because the two layers have different optical properties, it will induce non-smooth results of Stokes parameters distribution in the medium. In this sense, this case is more challenging for solving by the SEM than the above cases due to the spectral approximation tend to converge slowly for non-smooth result. The convergence characteristics of $h$- and $p$-refinement of the SEM for the solution of $G$ and $G_Q$ is presented in Fig. 9. The $h$-refinement test is conducted with a fixed polynomial order of $p = 2$, and the $p$-refinement test is conducted with a fixed number of elements $N_{el} = 2$. Though the convergence rate of $p$- refinement for solving $G$ and $G_Q$ is still faster than $h$-refinement, its convergence rate is much slower than for solving the above cases, which is due to the slow convergent problem of spectral approximation for non-smooth results. This illustrates that much caution should be paid for the SEM to solve the problems with non-smooth Stokes parameters distribution. For this kind of problems, low order spectral approximation with $h$-refinement is appropriate and a higher order spectral approximation might not improve much of accuracy.

### *3.4. Case 4: Two dimensional rectangular medium*

In this case, we consider radiative transfer in a two-dimensional oblique rectangular medium as shown in Fig. 10, the oblique angle to the x-direction is $\pi/3$. The optical height and optical width of the rectangle is 1 and 20, respectively. First, a problem that the rectangular medium is non-scattering medium is investigated. The temperature of the medium, the temperature of the lower boundary (z'=0) and the side boundaries are 300K. The top boundary is considered nonemitting. The lower boundary of the rectangle is a Fresnel reflection surface with refractive index of 3.724–2.212$i$. The refractive index of the medium and other boundary is 1. As the optical width of the rectangular is 20, which is large enough that the incident radiation from the side walls to the middle ($x' = 0$) of the rectangle is fully decayed and can be omitted. As such, the Stokes parameters along $z'$ at $x' = 0$ should be very close to the infinite flat layer problem. For the infinite layer problem, it is convenient to using ray tracing method to obtain a precise analytical solution (Appendix Section B), which can be taken as reference result. Here, The SEM is applied to solve the angular distributions of brightness temperature $T_{B,I}$ and $T_{B,Q}$ at frequency 85.5 Ghz. Figure 11(a) and 11(b) present the solved angular distributions of $T_{B,I}$ and $T_{B,Q}$, respectively, at three different positions, $z'/H$ = 0, 0.5, and 1. The results obtained with ray tracing technique for infinite layers are also shown as a comparison. In the solutions of the SEM, the angular discretization takes PCA schemes [18, 23] with $N_\theta \times N_\varphi = 40 \times 40$, and the rectangular medium is subdivided into 40 elements and 6$^{th}$ order polynomial approximation is used (as shown in Fig. 7, in which the dots denote the spectral nodes on each element). It is seen that the result obtained with the SEM agree very well with the result of ray tracing method for different component of polarization and at different position. The maximum relative error is less than 2%. This demonstrates the present SEM has good accuracy in solving two dimensional vector radiative transfer problem.

Second, a problem that the rectangular medium is scattering is investigated. An oblique solar beam (oblique angle respect to z' is $\cos\theta_c = 0.2$, and $\varphi_c = \pi/2$) penetrates into the medium similar to Case 2. The scattering phase matrix of the medium and other parameters is the same as that taken in Case 2. As the optical width of 20 is large enough that the incident radiation from the side walls to the middle ($x' = 0$) of the rectangle can be omitted, the Stokes parameters along $z'$ at $x' = 0$ should be very close to the infinite flat layer problem. Figure 12 presents the angular distributions of Stokes parameters at $z'/H$ =0.5 solved by the SEM with different order of spectral approximation for a scattering rectangular medium. The angular discretization takes PCA schemes with $N_\theta \times N_\varphi = 20 \times 40$, which has been demonstrated in Case 2 to be enough to obtain convergent results. The rectangular medium is subdivided into 40 elements and three different order of spectral approximation, namely, $p$ = 2, 4 and 6, are verified. It can be seen that the Stokes parameters obtained by the SEM gradually approach and finally agree well with the benchmark results of Garcia and Siewert [4] with increasing the order of spectral approximation, and the spectral approximation order of $p = 4$ is enough fine to obtain convergent results. This well illustrates the capability of the SEM to solve polarized radiative transfer in multidimensional scattering media.

## 4. Conclusions

A spectral element method is developed to solve polarized radiative transfer in multidimensional participating medium. The angular discretization is based on the discrete ordinates approach, and the spatial discretization is conducted by spectral element approach. Chebyshev polynomial is used to build basis function on each element. Four test problems are used to verify the performance of the SEM. The effectiveness of the SEM is demonstrated. The convergence rate of *p*-refinement follows the exponential rate of decrease and is superior to that of *h*-refinement. Higher order approximation is demonstrated to be accurate and efficient to solve the VRTE. The predicted angular distribution of brightness temperature and Stokes vector by the SEM agree very well with the benchmark solutions in references. Generally, the SEM is accurate, flexible, and effective to solve multidimensional polarized radiative transfer problems. However, it should be noted that the case in which the scattering phase function contains singularities is not considered in this paper. In practice, the problem of strong scattering anisotropy with sharp peak is very important [24]. Future work is still needed on extending the SEM to deal with this case.

**Acknowledgements**

The support of this work by the National Nature Science Foundation of China (50836002, 50620120442) is gratefully acknowledged.

**Appendix**

*A. Computation of $\cos\psi$ and $\cos\psi'$*

As shown in Fig. 1, the unit normal vectors of the scattering plane ($\mathbf{n}_{OPP'}$) and meridian planes of the incident ($\mathbf{n}_{OZP'}$) and scattered radiation ($\mathbf{n}_{OZP}$) can be obtained as

$$\mathbf{n}_{OPP'} = \frac{\mathbf{\Omega}\times\mathbf{\Omega}'}{|\mathbf{\Omega}\times\mathbf{\Omega}'|}, \quad \mathbf{n}_{OZP'} = \frac{\mathbf{k}\times\mathbf{\Omega}'}{|\mathbf{k}\times\mathbf{\Omega}'|}, \quad \mathbf{n}_{OZP} = \frac{\mathbf{k}\times\mathbf{\Omega}}{|\mathbf{k}\times\mathbf{\Omega}|} \tag{20}$$

where $\mathbf{k}$ is the unit vector of z direction. With the help of these vectors [Eq.(20)], $\cos\psi'$ and $\cos\psi$ are convenient to be expressed as

$$\cos\psi' = \mathbf{n}_{OZP'}\bullet\mathbf{n}_{OPP'}, \quad \cos\psi = -\mathbf{n}_{OZP}\bullet\mathbf{n}_{OPP'}, \text{ for } \mathbf{n}_{OPP'}\bullet\mathbf{k} \geq 0 \tag{21a}$$

$$\cos\psi' = -\mathbf{n}_{OZP'}\bullet\mathbf{n}_{OPP'}, \quad \cos\psi = \mathbf{n}_{OZP}\bullet\mathbf{n}_{OPP'}, \text{ for } \mathbf{n}_{OPP'}\bullet\mathbf{k} < 0 \tag{21b}$$

If $\mathbf{n}_{OPP'}\bullet\mathbf{k} = 0$ or $\mathbf{\Omega}'\times\mathbf{\Omega} = 0$, the meridian planes of the two beam coincides and which aligns with the scattering plane, thus no rotation is need, namely, $\psi' = 0$ and $\psi = 0$, or, $\cos\psi' = 1$ and $\cos\psi = 1$, which is a result of Eq. (21a). In the condition of Eq. (21b), it is the case of $\cos\psi' = -1$ and $\cos\psi = -1$, which correspond to a rotation angle of $\pi$. As seen from Eq. (3), the rotation angles of 0 and $\pi$ are equivalent for the rotation matrix.

*B. Ray tracing between infinite homogeneous nonscattering slab with considering the effect of polarization*

The configuration of the slab is given in Fig. 13. The height of the slab is $H$, and the medium in the slab is homogenous and nonscattering. The extinction matrix of the medium is diagonal with scalar extinction coefficient $\kappa_e$. The reflection matrix of the lower and top wall is $\overline{\overline{\mathbf{R}}}_0$ and $\overline{\overline{\mathbf{R}}}_1$, respectively, which are functions of the incident angle $\theta$, the medium refractive index and the outer space refractive index, and are computed from Eq.

(5). We first consider the case that the medium is nonemitting, while the lower and the top walls are emitting with Stokes vector of $\mathbf{I}_0$ and $\mathbf{I}_1$, respectively.

As shown in Fig. 13, by using reverse ray tracing, the Stokes vector $\mathbf{I}_W(z,\theta)$ of point $P$ is contributed from all the emitting source point on the wall, as denoted as 0, 1, 2, …,$k$, which denotes the $k$-th contribution. For zenith angle $\theta \in [0, \pi/2]$, the contribution of the zeroth emitting source point $\mathbf{C}_0$ can be computed as

$$\mathbf{C}_0 = \mathbf{I}_0 e^{-\kappa_e d_0} \tag{22}$$

where $d_0 = z/\cos\theta$ is the distance between point $P$ and the zeroth contribution point. The contribution of each even number ($k = 2l$) emitting source point $\mathbf{C}_k$ can be computed as

$$\mathbf{C}_k = (\overline{\overline{\mathbf{R}}}_0 \overline{\overline{\mathbf{R}}}_1)^l \mathbf{I}_0 e^{-\beta d_k}, \qquad l = 0,1,2... \tag{23}$$

where $d_k = kH/\cos\theta + d_0$ is the distance between point $P$ and the $k$-th contribution point along the ray trajectory. The contribution of each odd number ($k = 2l+1$) emitting source point $\mathbf{C}_k$ can be computed as

$$\mathbf{C}_k = (\overline{\overline{\mathbf{R}}}_0 \overline{\overline{\mathbf{R}}}_1)^l \overline{\overline{\mathbf{R}}}_0 \mathbf{I}_1 e^{-\beta d_k}, \qquad l = 0,1,2... \tag{24}$$

The Stokes vector $\mathbf{I}_W(z,\theta)$ ($\theta \in [0, \pi/2]$) can be obtained by summation of all the contributions, it follows

$$\begin{aligned}
\mathbf{I}_W(z,\theta) &= \sum_{k=0}^{\infty} \mathbf{C}_k = \sum_{l=0}^{\infty} \left\{ (\overline{\overline{\mathbf{R}}}_0 \overline{\overline{\mathbf{R}}}_1)^l \mathbf{I}_0 e^{-\kappa_e d_{2l}} + (\overline{\overline{\mathbf{R}}}_0 \overline{\overline{\mathbf{R}}}_1)^l \overline{\overline{\mathbf{R}}}_0 \mathbf{I}_1 e^{-\kappa_e d_{2l+1}} \right\} \\
&= \sum_{l=0}^{\infty} \left\{ (\overline{\overline{\mathbf{R}}}_0 \overline{\overline{\mathbf{R}}}_1 e^{-2\kappa_e H/\cos\theta})^l \mathbf{I}_0 e^{-\kappa_e d_0} + (\overline{\overline{\mathbf{R}}}_0 \overline{\overline{\mathbf{R}}}_1 e^{-2\kappa_e H/\cos\theta})^l \overline{\overline{\mathbf{R}}}_0 \mathbf{I}_1 e^{-\kappa_e (H/\cos\theta + d_0)} \right\} \\
&= \left[ \overline{\overline{\mathbf{E}}} - \overline{\overline{\mathbf{R}}}_0 \overline{\overline{\mathbf{R}}}_1 e^{-2\kappa_e H/\cos\theta} \right]^{-1} \left[ \mathbf{I}_0 + \overline{\overline{\mathbf{R}}}_0 \mathbf{I}_1 e^{-\kappa_e H/\cos\theta} \right] e^{-\kappa_e d_0}
\end{aligned} \tag{25}$$

When $\theta \in [\pi/2, \pi]$, the Stokes vector can be obtained in a similar analysis as

$$\mathbf{I}_W(z,\theta) = \left[ \overline{\overline{\mathbf{E}}} - \overline{\overline{\mathbf{R}}}_1 \overline{\overline{\mathbf{R}}}_0 e^{-2\kappa_e H/|\cos\theta|} \right]^{-1} \left[ \mathbf{I}_1 + \overline{\overline{\mathbf{R}}}_1 \mathbf{I}_0 e^{-\kappa_e H/|\cos\theta|} \right] e^{-\kappa_e \tilde{d}_0} \tag{26}$$

Where $\tilde{d}_0 = (H-z)/|\cos\theta|$.

The derivation presented above assumes the medium is nonemitting. In the following, the medium is uniformly emitting with Stokes vector $\mathbf{I}_b$. Here, we consider the case when the boundary is nonemitting. For the case the boundary and the medium are all emitting, it can be easily dealt with by superposition principle.

For zenith angle $\theta \in [0, \pi/2]$, as shown in Fig. 13, by using an equivalent approach, the contribution of the emitting from medium between the zeroth point and point $P$ can be taken as a point source just like emitting from the wall, it follows the zeroth contribution to be calculated as

$$\mathbf{C}_0 = \mathbf{I}_{g0} e^{-\kappa_e d_0} \tag{27}$$

where $\mathbf{I}_{g0} = \mathbf{I}_b \left[ \exp(\kappa_e d_0) - 1 \right]$ is the equivalent emitting to the zeroth point. With this equivalent approach, the medium contributions are taken as emission from the wall, which is similar to the previous case discussed. In a similar manner, the contribution of each even number ($k = 2l$) equivalent emitting source point $\mathbf{C}_k$ (contribution of emitting from medium between points $k-1$ and $k$) can be computed as

$$\mathbf{C}_k = (\bar{\bar{\mathbf{R}}}_0 \bar{\bar{\mathbf{R}}}_1)^l \mathbf{I}_g e^{-\kappa_e d_k}, \qquad l = 1, 2... \tag{28}$$

where $\mathbf{I}_g = \mathbf{I}_b \left[ \exp(\kappa_e H / \cos\theta) - 1 \right]$. The contribution of each odd number ($k = 2l+1$) equivalent emitting source point $\mathbf{C}_k$ (contribution of emitting from medium between points $k-1$ and $k$) can be computed as

$$\mathbf{C}_k = (\bar{\bar{\mathbf{R}}}_0 \bar{\bar{\mathbf{R}}}_1)^l \bar{\bar{\mathbf{R}}}_0 \mathbf{I}_g e^{-\kappa_e d_k}, \qquad l = 0, 1, 2... \tag{29}$$

The Stokes vector $\mathbf{I}_M(z, \theta)$ ($\theta \in [0, \pi/2]$) can be obtained by summation of all the contributions, it follows

$$\begin{aligned} \mathbf{I}_M(z, \theta) &= \sum_{k=0}^{\infty} \mathbf{C}_k = \mathbf{I}_{g0} e^{-\kappa_e d_0} + \sum_{l=1}^{\infty} (\bar{\bar{\mathbf{R}}}_0 \bar{\bar{\mathbf{R}}}_1)^l \mathbf{I}_g e^{-\kappa_e d_{2l}} + \sum_{l=0}^{\infty} (\bar{\bar{\mathbf{R}}}_0 \bar{\bar{\mathbf{R}}}_1)^l \bar{\bar{\mathbf{R}}}_0 \mathbf{I}_g e^{-\kappa_e d_{2l+1}} \\ &= \mathbf{I}_{g0} e^{-\kappa_e d_0} - \mathbf{I}_g e^{-\kappa_e d_0} + \sum_{l=0}^{\infty} \left\{ (\bar{\bar{\mathbf{R}}}_0 \bar{\bar{\mathbf{R}}}_1 e^{-2\kappa_e H/\cos\theta})^l \mathbf{I}_g e^{-\kappa_e d_0} + (\bar{\bar{\mathbf{R}}}_0 \bar{\bar{\mathbf{R}}}_1 e^{-2\kappa_e H/\cos\theta})^l \bar{\bar{\mathbf{R}}}_0 \mathbf{I}_g e^{-\kappa_e (H/\cos\theta + d_0)} \right\} \\ &= (\mathbf{I}_{g0} - \mathbf{I}_g) e^{-\kappa_e d_0} + \left[ \bar{\bar{\mathbf{E}}} - \bar{\bar{\mathbf{R}}}_0 \bar{\bar{\mathbf{R}}}_1 e^{-2\kappa_e H/\cos\theta} \right]^{-1} \left[ \mathbf{I}_g + \bar{\bar{\mathbf{R}}}_0 \mathbf{I}_g e^{-\kappa_e H/\cos\theta} \right] e^{-\kappa_e d_0} \end{aligned} \tag{30}$$

When $\theta \in [\pi/2, \pi]$, the Stokes vector can be obtained in a similar analysis as

$$\mathbf{I}_M(z, \theta) = (\mathbf{I}_{g0} - \mathbf{I}_g) e^{-\kappa_e \tilde{d}_0} + \left[ \bar{\bar{\mathbf{E}}} - \bar{\bar{\mathbf{R}}}_1 \bar{\bar{\mathbf{R}}}_0 e^{-2\kappa_e H/|\cos\theta|} \right]^{-1} \left[ \mathbf{I}_g + \bar{\bar{\mathbf{R}}}_1 \mathbf{I}_g e^{-\kappa_e H/|\cos\theta|} \right] e^{-\kappa_e \tilde{d}_0} \tag{31}$$

If the medium and the wall are both emitting, then the total Stokes vector $\mathbf{I}(z, \theta)$ is obtained from superposition principle as

$$\mathbf{I}(z, \theta) = \mathbf{I}_W(z, \theta) + \mathbf{I}_M(z, \theta) \tag{32}$$

**Figure Captions**

**Fig. 1.** Geometry for the definition of polarization reference frame and definition of angles in the transformation of scattering phase matrix.

**Fig. 2**. Angular distributions of brightness temperature at different positions solved by the SEM for Case 1: **(a)** $T_{B,I}$ and **(b)** $T_{B,Q}$.

**Fig. 3.** The convergence characteristics of h- and p- refinement of the SEM at different optical thickness for solving: **(a)** $G$ and **(b)** $G_Q$.

**Fig. 4.** The convergence characteristics of h-refinement under different order of spectral approximation.

**Fig. 5.** Schematic of the oblique incident solar beam irradiate into a layer of atmosphere.

**Fig. 6.** Angular distribution of the four Stokes parameters solved by the SEM at three different positions for Case 2: **(a)** $z/H$ =0, **(b)** $z/H$ =0.5 and **(c)** $z/H$ =1.

**Fig. 7.** The convergence characteristics of h- and p- refinement of the SEM for solving Case 2.

**Fig. 8.** Angular distribution of brightness temperatures at the top and bottom of the atmosphere solved by the SEM for Case 3: **(a)** $T_{B,I}$ and **(b)** $T_{B,Q}$.

**Fig. 9.** The convergence characteristics of h- and p- refinement of the SEM for solving Case 3.

**Fig. 10.** Schematic of the two-dimensional oblique rectangular medium.

**Fig. 11.** Angular distributions of brightness temperature at different positions solved by the SEM for a nonscattering rectangular medium: **(a)** $T_{B,I}$ and **(b)** $T_{B,Q}$.

**Fig. 12.** Angular distributions of Stokes parameters at $z'/H$ =0.5 solved by the SEM with different order of spectral approximation for a scattering rectangular medium.

**Fig. 13.** Schematic of reverse ray tracing in one dimensional infinite slab.

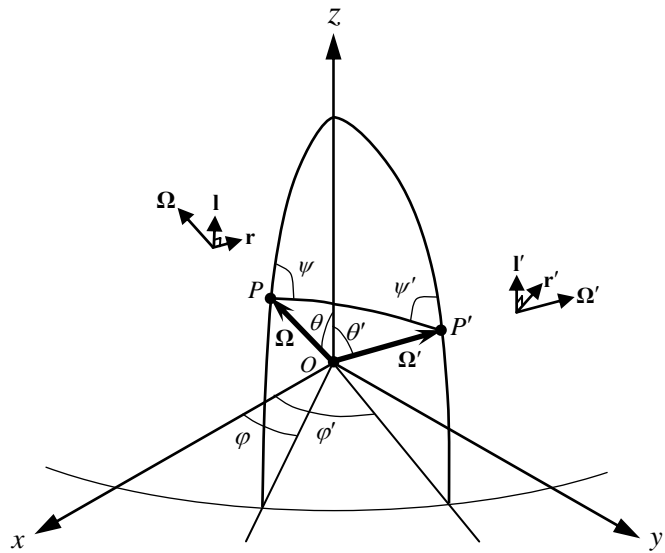

**Figure 1**

Authors: Zhao, Liu, Hsu and Tan

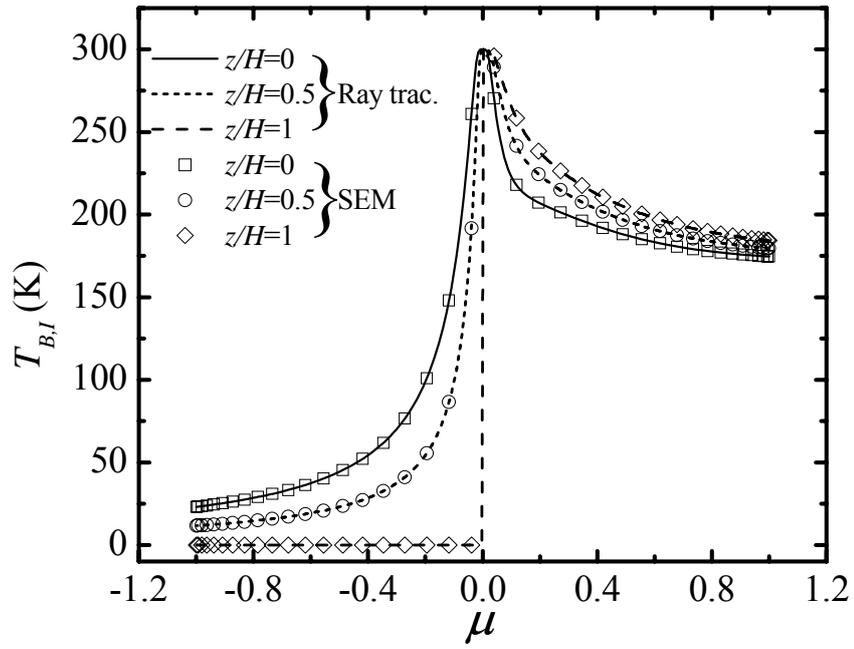

**Figure 2(a)**

**Authors: Zhao, Liu, Hsu and Tan**

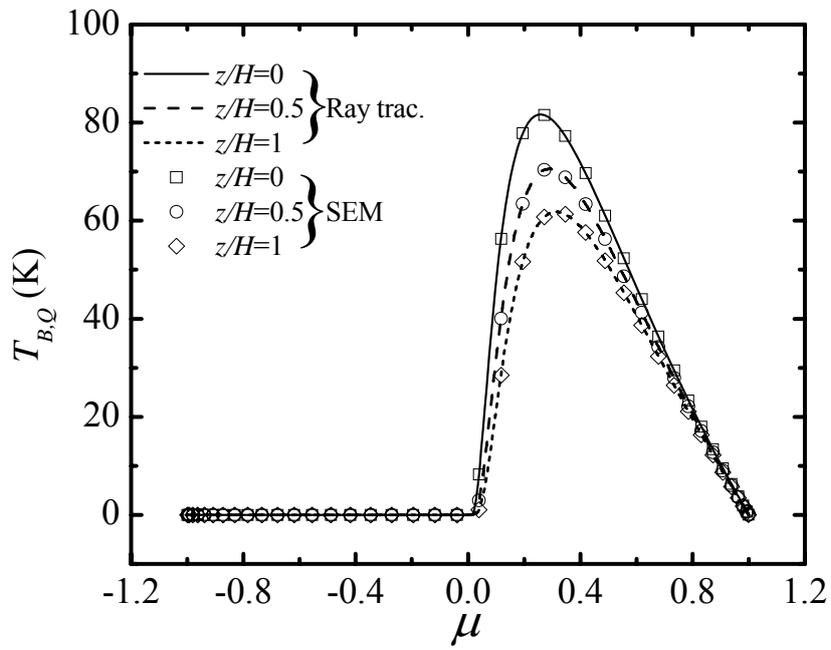

**Figure 2(b)**

**Authors: Zhao, Liu, Hsu and Tan**

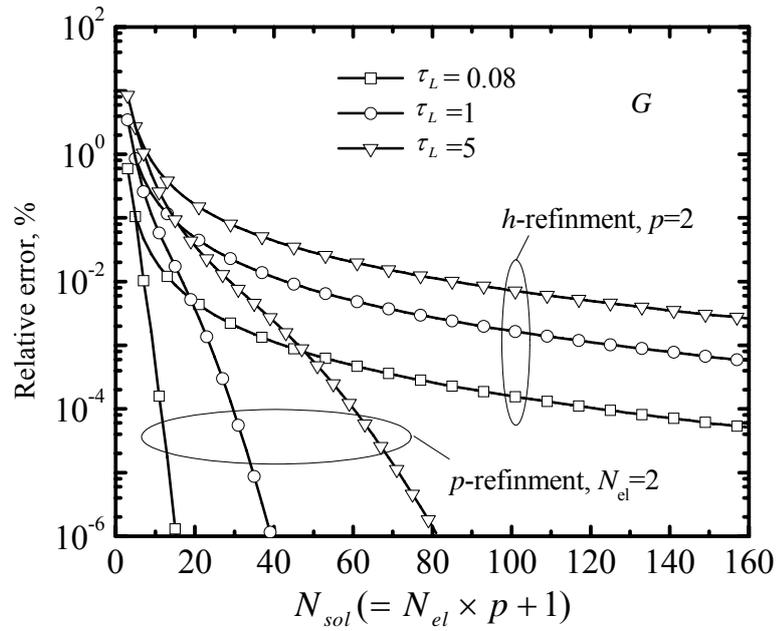

**Figure 3(a)**

**Authors: Zhao, Liu, Hsu and Tan**

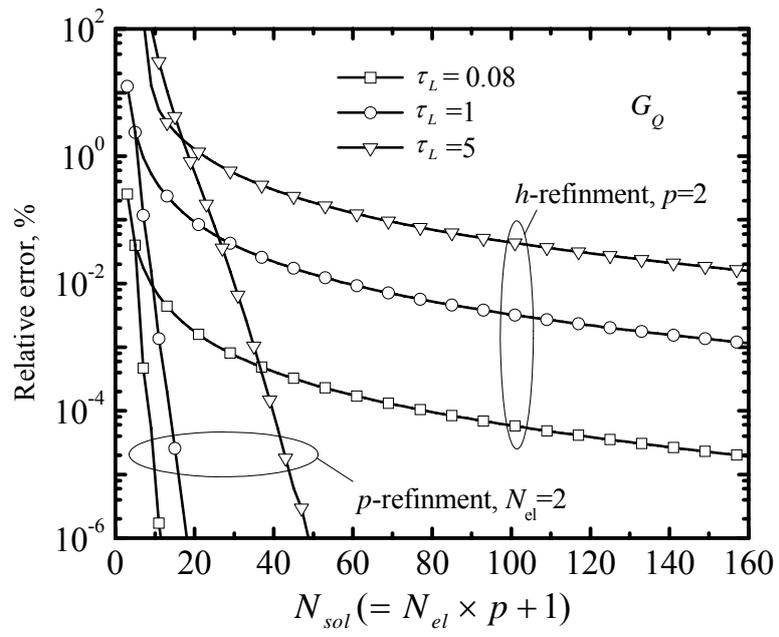

**Figure 3(b)**

**Authors: Zhao, Liu, Hsu and Tan**

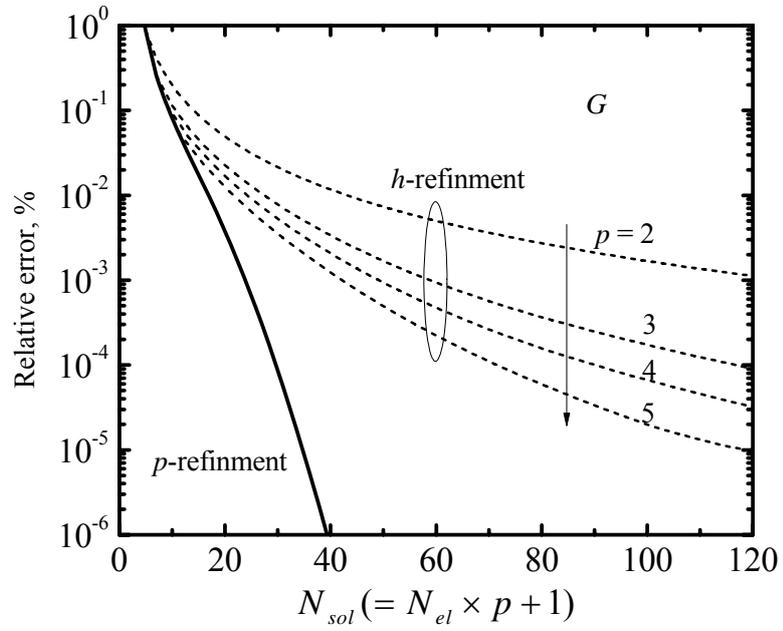

**Figure 4**

Authors: Zhao, Liu, Hsu and Tan

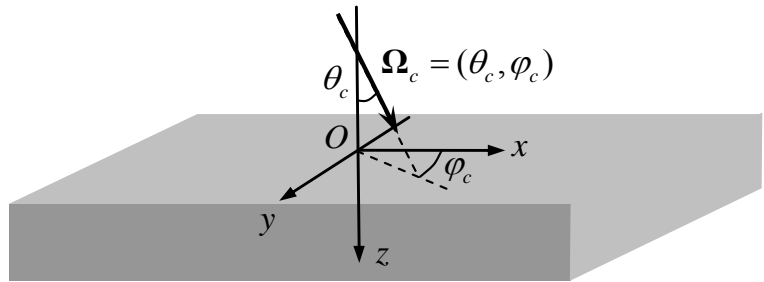

**Figure 5**

**Authors: Zhao, Liu, Hsu and Tan**

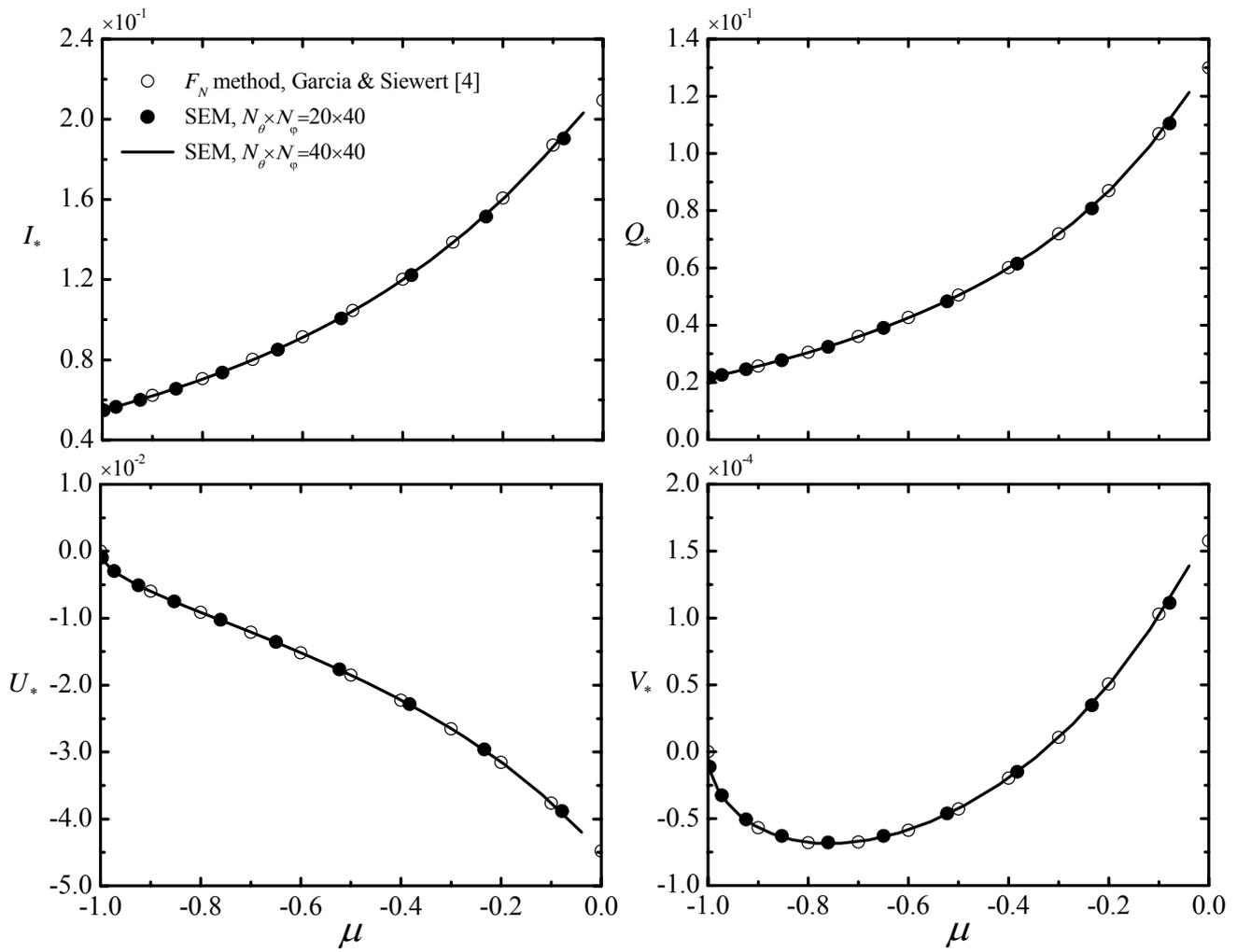

**Figure 6(a)**

**Authors: Zhao, Liu, Hsu and Tan**

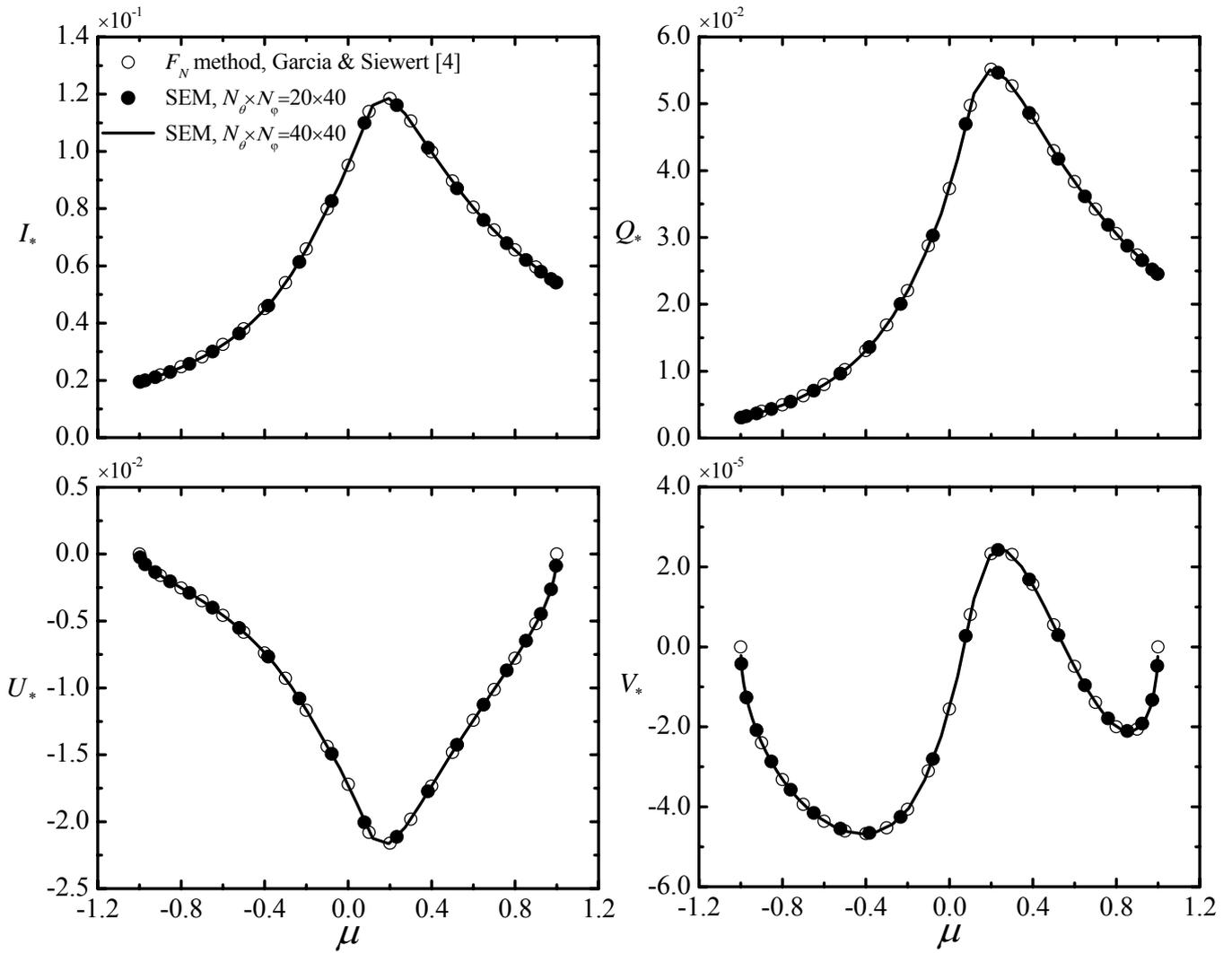

**Figure 6(b)**

**Authors: Zhao, Liu, Hsu and Tan**

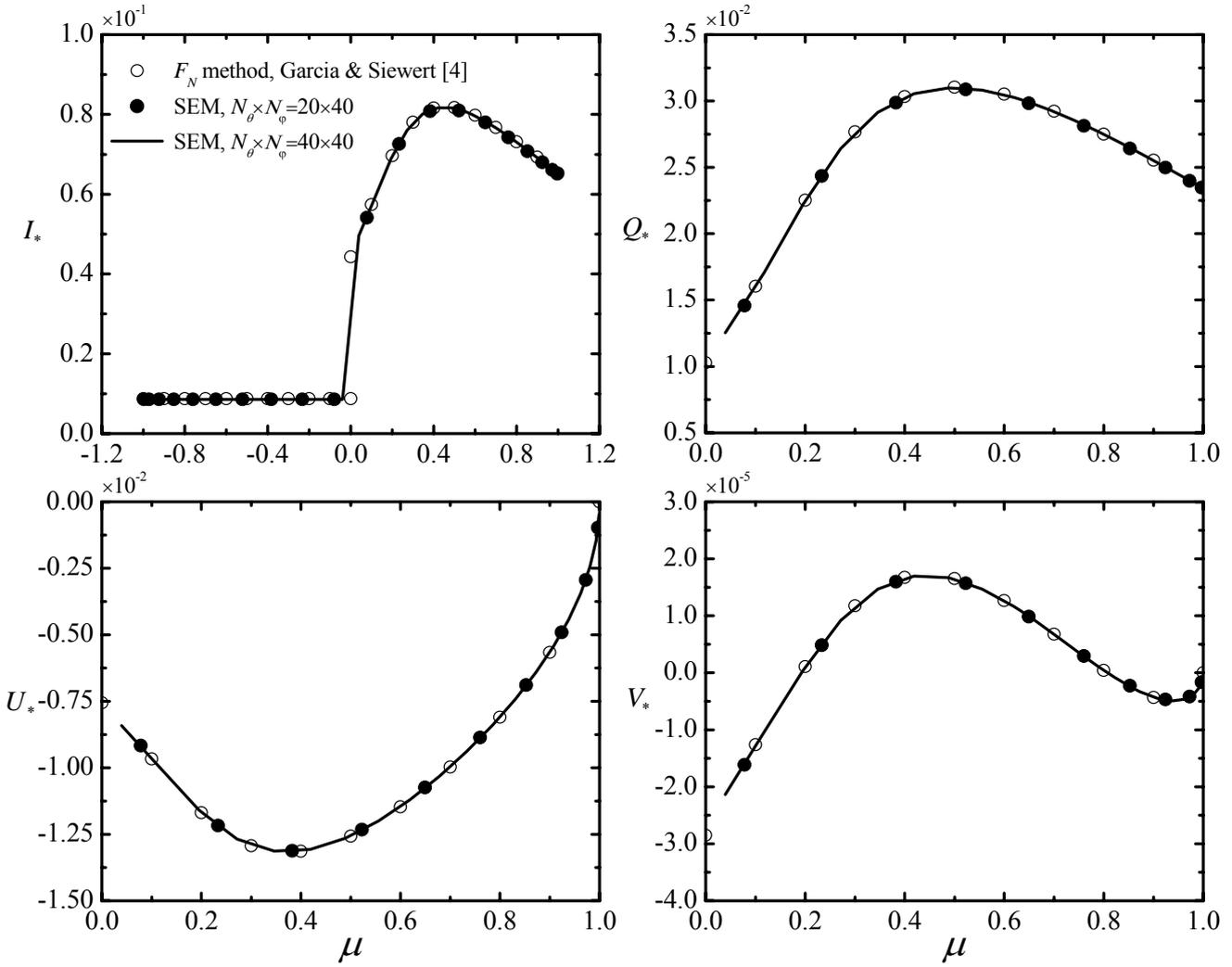

**Figure 6(c)**

**Authors: Zhao, Liu, Hsu and Tan**

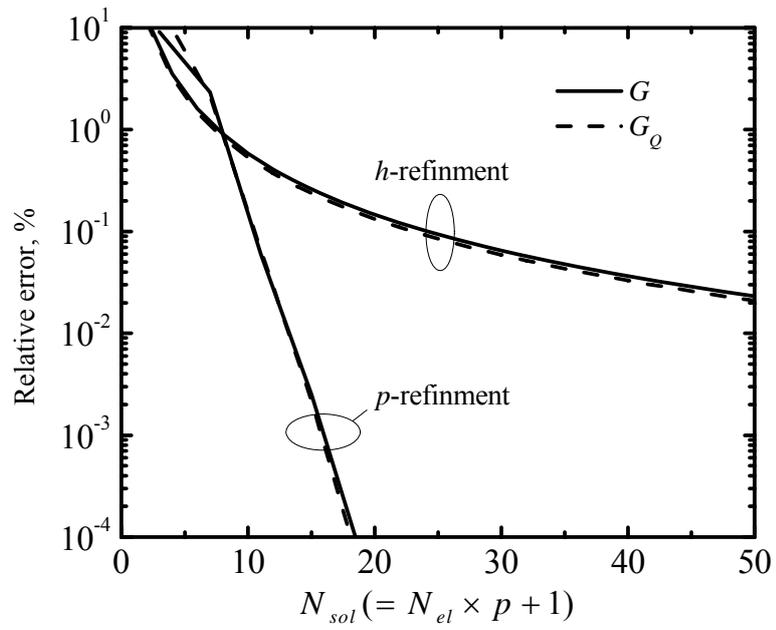

**Figure 7**

**Authors: Zhao, Liu, Hsu and Tan**

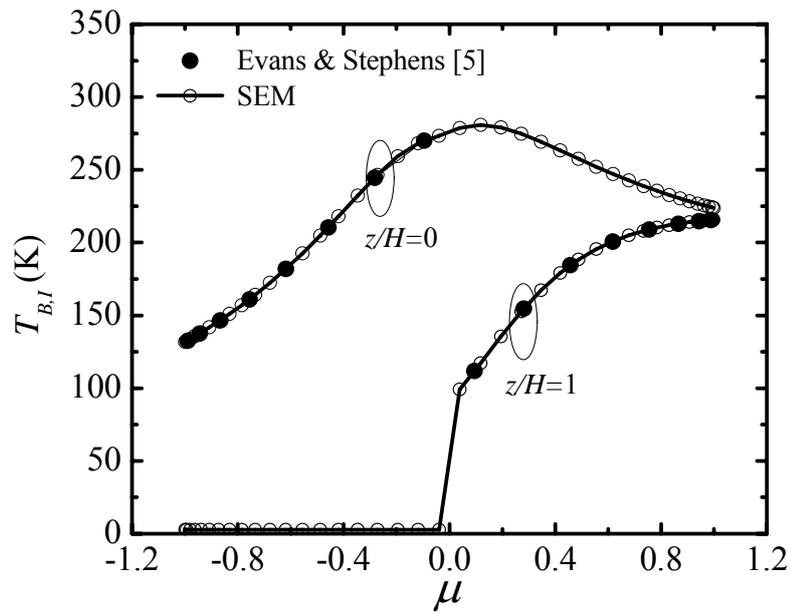

**Figure 8(a)**

**Authors: Zhao, Liu, Hsu and Tan**

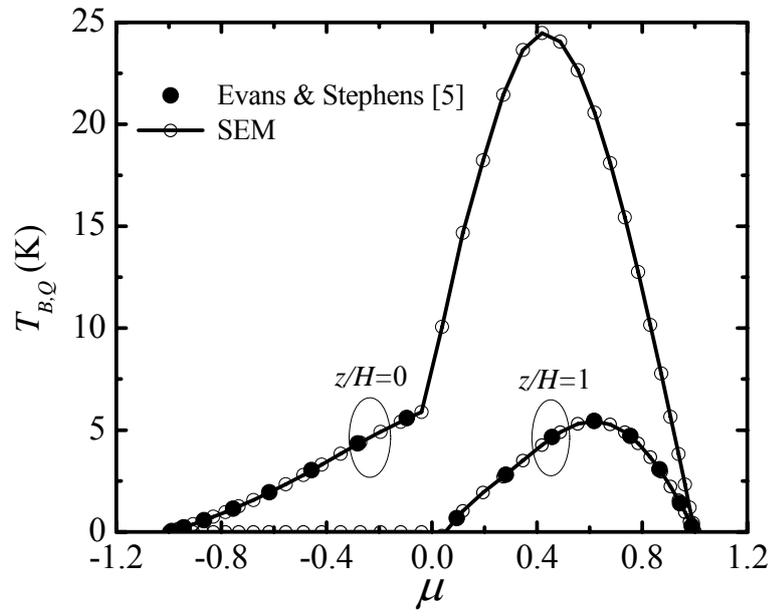

**Figure 8(b)**

**Authors: Zhao, Liu, Hsu and Tan**

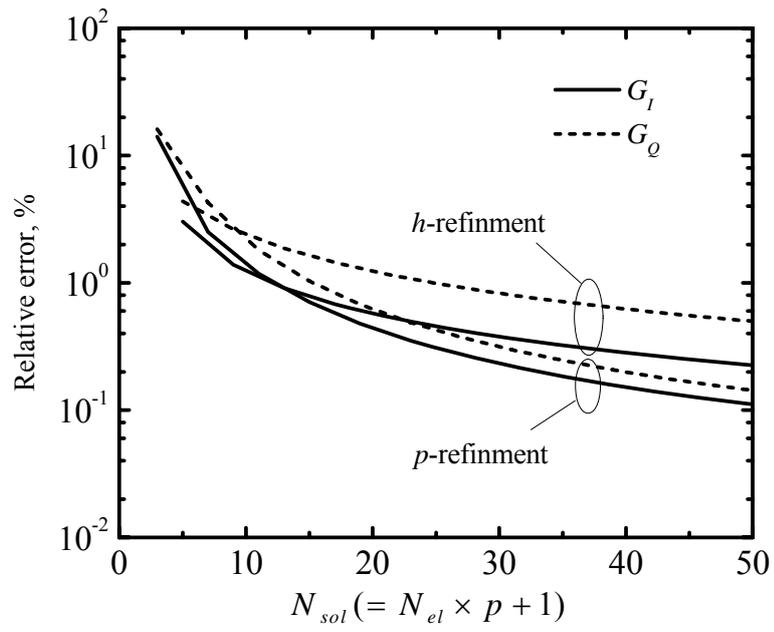

**Figure 9**

**Authors: Zhao, Liu, Hsu and Tan**

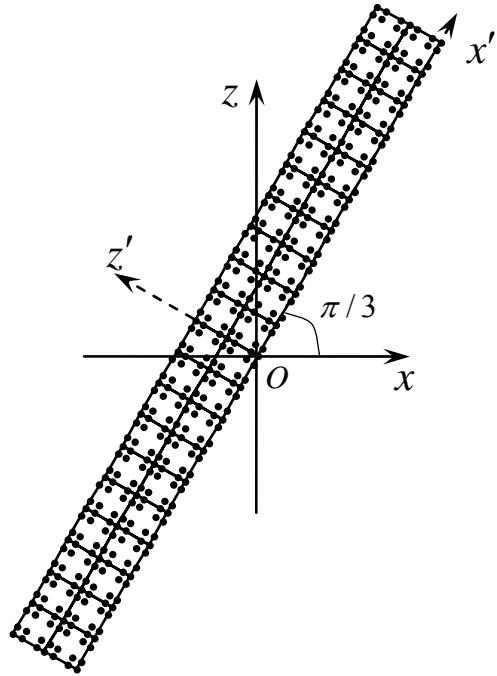

**Figure 10**

**Authors: Zhao, Liu, Hsu and Tan**

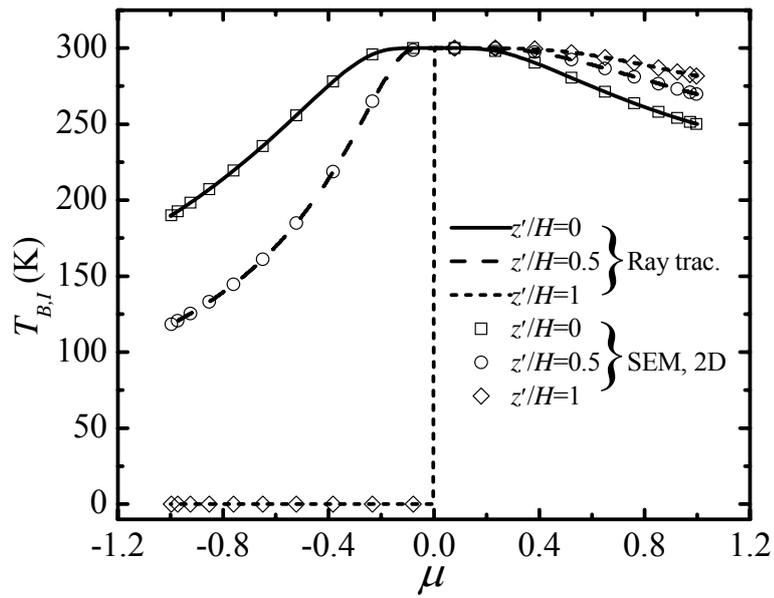

**Figure 11(a)**

**Authors: Zhao, Liu, Hsu and Tan**

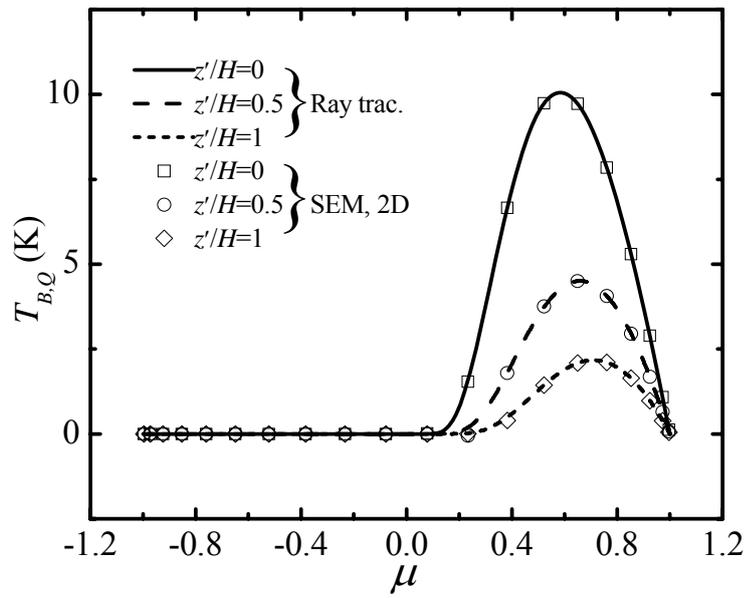

**Figure 11(b)**

Authors: Zhao, Liu, Hsu and Tan

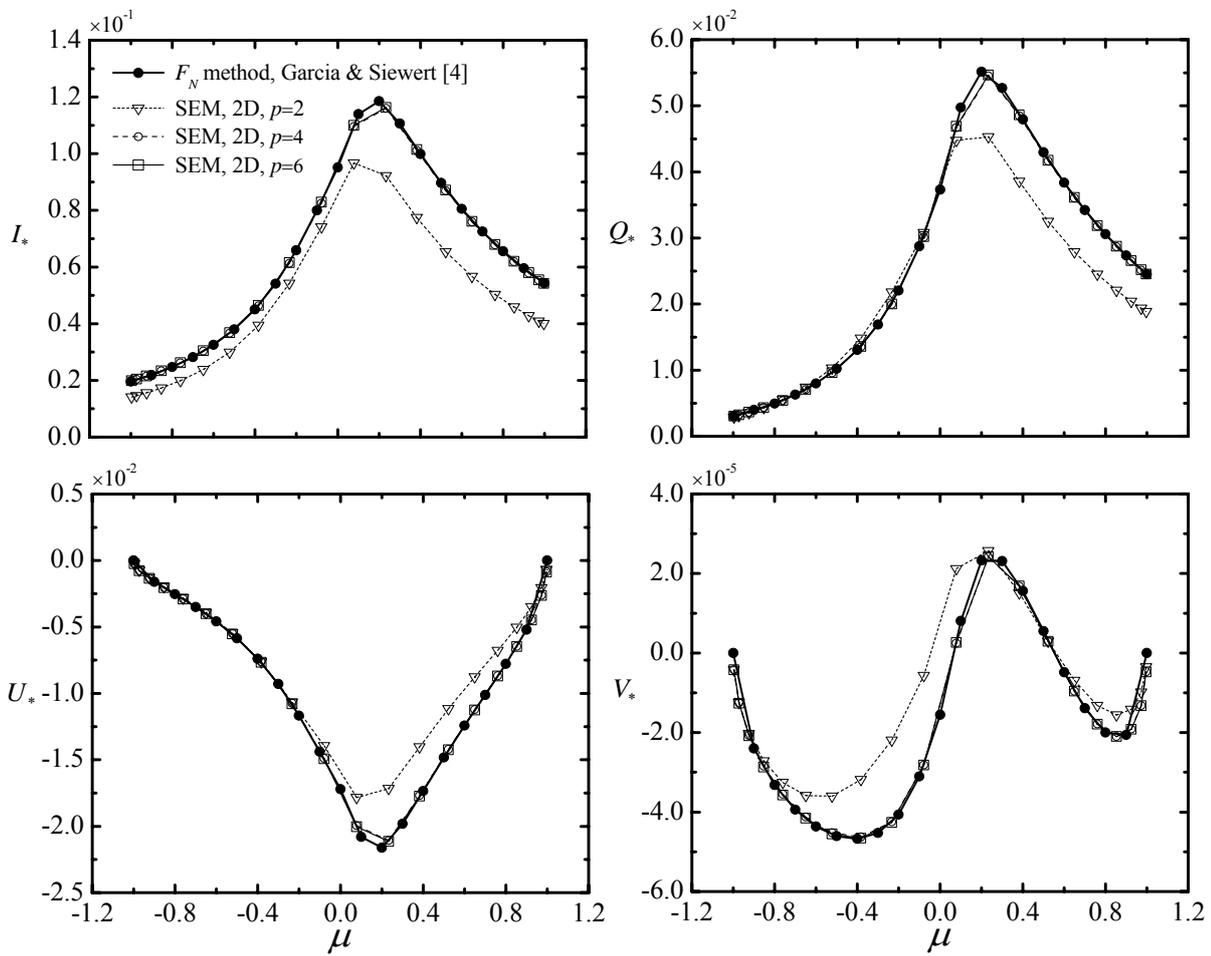

**Figure 12**

Authors: Zhao, Liu, Hsu and Tan

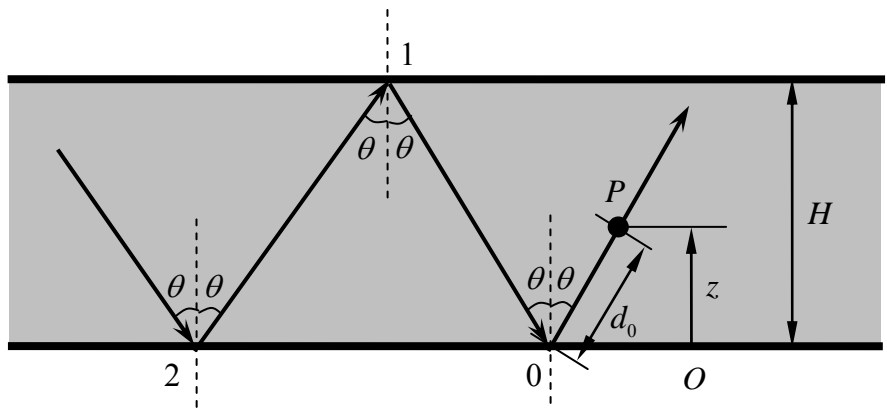

**Figure 13**

**Authors: Zhao, Liu, Hsu and Tan**